# Ethical behavior in humans and machines
## Evaluating training data quality for beneficial machine learning

Dr. Thilo Hagendorff
thilo.hagendorff@uni-tuebingen.de
University of Tuebingen
Cluster of Excellence "Machine Learning: New Perspectives for Science"
International Center for Ethics in the Sciences and Humanities

**Abstract** – Machine behavior that is based on learning algorithms can be significantly influenced by the exposure to data of different qualities. Up to now, those qualities are solely measured in technical terms, but not in ethical ones, despite the significant role of training and annotation data in supervised machine learning. This is the first study to fill this gap by describing new dimensions of data quality for supervised machine learning applications. Based on the rationale that different social and psychological backgrounds of individuals correlate in practice with different modes of human-computer-interaction, the paper describes from an ethical perspective how varying qualities of behavioral data that individuals leave behind while using digital technologies have socially relevant ramification for the development of machine learning applications. The specific objective of this study is to describe how training data can be selected according to ethical assessments of the behavior it originates from, establishing an innovative filter regime to transition from the big data rationale *n = all* to a more selective way of processing data for training sets in machine learning. The overarching aim of this research is to promote methods for achieving beneficial machine learning applications that could be widely useful for industry as well as academia.

## 1  Introduction

In a recent seminal paper on machine behavior, Rahwan et al. (2019) stress that machine learning applications "cannot be fully understood without the integrated study of algorithms and the social environments in which algorithms operate". With regard to supervised machine learning, those "social environments" can, among others, be understood as different training stimuli that shape the behavior of a machine. Machine behavior can be seen in analogy to the behavior of biological agents as an observable response to (internal or) external stimuli. Training data fed into machine learning applications reflect, in case it is about behavioral data, people's (e.g., discriminative) behavior, so people's behavior becomes machine (discriminative) behavior (Barocas and Selbst 2016). Thus, when technology ethicists talk about "moral machines" (Wallach and Allen 2009) in the context of machine learning applications, one also has to ask for "moral people" and "moral people's data", to put it simply. Of course, these "moral machines" are also the result of engineering or design choices, they are dependent on the selection of hyperparameters or specific wirings of artificial neural networks, and the like. But in general, today's machine learning techniques are dependent on human participation. In many cases, they harness human behavior that is digitized by various tracking methods. These machine learning methods do not create intelligence, but capture it by tracking human cognitive and behavioral abilities. Without the empirical aggregation of recordings of human behavior, many parts of machine learning would not work. An extensive infrastructure for capturing human behavior in distributed networks builds the bedrock for a computational capacity called "artificial intelligence" (Mühlhoff 2019).

Here, I want to ask whether there are differences in the "quality" of human participation in artificial intelligence. To do this, one must further answer the question about what constitutes "good" influences or "good" behavioral datasets for machine learning applications. In order to accomplish this, I will focus on identifying data sources reflecting behavior that is ethically sound, which in turn can be identified via scrutinizing particular states and traits of an individual that are to be described in more detail. With the help of a matrix of different evaluation frameworks, a normative evaluation of different data sources can take place. To the best of my knowledge, such an approach has not yet been enlarged upon in the computer sciences. Hitherto, normatively oriented machine learning research is mainly concerned with fairness (Kearns and Roth 2020) or preventing discrimination (Hagendorff 2019c), robustness (Amodei et al. 2017), explainability (Mittelstadt et al. 2019), or preserving privacy (Dwork 2006). Besides that, especially in the field of supervised machine learning, the question of what characterizes – from an ethical perspective – good data contexts remains largely unanswered. This is crucial, since morally sound machine learning applications are in many regards only as sophisticated as their "environmental influences" or training stimuli.

Fruitful research can emerge when the social sciences are combined with machine learning research, so that not only ethics, but also technology development can be advanced. Most research works in this area provide critique rather than engage constructively by creating positive ideas and visions on how to use machine learning technologies for the common good. This paper stands in line with and continues the "good data project" (Daly et al. 2019), promoting tangible good and ethical data practices and frameworks instead of just criticizing what goes wrong with machine learning and big data applications (Crawford et al. 2019; Whittaker et al. 2018; Campolo et al. 2017; Crawford et al. 2016). The following chapters elaborate on that in more detail. Chapter 2.1 describes the present approach which is to use as much behavioral data as possible for machine learning development. Although criteria for data quality exist to prefilter training stimuli, these criteria are solely oriented along technical dimensions, not ethical ones, as depicted in chapter 2.2. Chapter 3.1 then describes how human behavior is normatively classified in sociology and psychology, while chapter 3.2 describes how tracking technologies can be inspired by those classifications in order to single out datasets from certain subpopulations that are deemed to be the most competent or morally versed group for a particular task. Chapter 4 then investigates four particular applications for machine learning, namely autonomous cars, language generation, social media filtering systems, search engine ranking algorithms, and e-commerce recommendation systems, that can be made more beneficial by following the presented ideas. To that end, machine behavior objectives, behavioral data sources, tracked states and traits to assess the quality of those data, as well as quality training stimuli of each of the example applications are to be described. Subsequently, chapter 5 covers some points of discussion and responds to them defending the presented approach for beneficial machine learning. Finally, chapter 6 concludes and sums up the paper's arguments.

## 2 More data are not always better: defining new dimensions of data quality

### 2.1 The idea of *n = all* and its shortcomings

Before deliberating on data quality dimensions, I want to recapitulate the tenets of big data. Big data meant the emancipation from small data studies, a paradigm scientific knowledge discovery relied on for hundreds of years. Big data lead to the success of today's machine learning systems, which are heralded as the new gold standard of knowledge discovery since they are necessary to understand increasingly complex collections of data, especially in the sciences (Mjolsness and DeCoste 2001; Jordan and Mitchell 2015). Broadly speaking, this trend caused some kind of "amnesia" on the value of small data (Kitchin and Lauriault 2015). While small data are of narrow variety, have a limited volume, are generated to answer specific questions, and produced in controlled ways, big data are the exact opposite. The latter are large in volume since they are generated continuously as a by-product of digital technologies. As stated many times, big data strives to be exhaustive in scope, or, in other words, it follows the ideology of *n = all*. The formula *n = all* encapsulates the idea that "more trumps better" (Mayer-Schönberger and Cukier 2013, p. 13; Perrons and McAuley 2015). Hence, big data are often indifferent towards predefined, specific queries or areas of interest in the context of which one wants to gather insights. Queries often repurpose data to gain insights into phenomena that have no or only indirect linkage to the original context of the data acquisition.

Machine learning techniques allow probabilistic inferences on unknown features. This is why current machine learning applications work under the motto "the more data they have, the better they get" (Domingos 2015, p. xi). But when speaking about behavioral data, this claim may not be true. It seems that learning applications don't have to be programmed, they program themselves. But when they program themselves while being fed with as much behavioral data as possible in order to aim at higher grades of accuracy, they also become indifferent with respect to the orientation towards certain moral values. The ideology of *n = all* leads to technical systems that utilize an endless stream of choices made by humans interacting with online platforms and digital devices – a practice once called "laissez-faire data collection" (Jo and Gebru 2019) –, narrowing down everything towards scores which represent averages of whole populations. But instead of simply recognizing patterns within datasets of a whole population, one could single out datasets – and hence training stimuli – from a certain subpopulation, namely the most competent, eligible, or morally versed one for a respective task, and find patterns only within this data context.

Subsequently, only those patterns provide the basis for the generalization ability of a given model or learning algorithm. By diversifying or sampling various data contexts in the larger frame of big data, one can reintroduce ideas connected to the concept of small data in the current situation of an abundance of data. This abundance is so prevalent that measures to tailor training data sets for machine learning applications with respect to certain fractions of large data sets does not mean to significantly restrict technical capabilities of these applications. But the decisive question is what fractions to choose, what data nuances to accentuate. In this context, the claim that "more trumps better" is transposed to "better trumps more". But what is better data? To answer this question, one has to look at data quality.

## 2.2 Data quality dimensions and ethics

A common saying in computer sciences is "garbage in, garbage out", referring to the importance of quality data in data-intensive applications like supervised machine learning. Surveys found that data quality attributes comprise literally hundreds of variables (Wang and Strong 1996). Nevertheless, discourses on data quality define it in terms of its suitability for a business purpose and decision-making efficiency in companies (Samitsch 2015), and are solely focused on particular technical dimensions like data cleanliness (how many errors do data sets contain?), data completeness (how exhaustive for a particular task are data sets?), data objectivity (what biases do data sets contain?), data consistency and reliability (how many discrepancies are contained in data sets?), data timeliness (how current are the data?), data veracity and exactitude (how accurate and precise are information in data sets?), data interpretability (how readable are data sets?), data cost-effectiveness (what are the costs of data collections?), and the like (Gudivada et al. 2017). Similarly, data quality problems are defined in terms of missing data, duplicate data, inconsistent data formats, incorrect values, spelling errors, etc. (Woodall et al. 2014). Current approaches to improve datasheets for datasets also do not include aspects that go beyond technical and organizational items like "Who funded the creation of the datasets?", "Are there any errors, sources of noise, or redundancies in the dataset?", "Will the datasets be updated?", and so on (Gebru et al. 2018). All those questions and differentiations make perfect sense when assessing data sets that do not contain data that relate to human behavior. But in case data sets relate to it, the discourse on data quality has to be extended.

Besides the data quality discourse, a further discourse addresses the construction of digital persons via data traces from volatile and non-volatile data acquisitions, sensors of all kinds, surveillance measures, social media platforms, and the like. Personal data from different sources and domains are linked together in a form of a "dense rhizomatic assemblage" (Kitchin and Dodge 2011, p. 90). Terms like "data subjects", "data derivatives" (Amoore 2011), "data double" (Los 2006), "shadow order" (Bogard 1996), "digital persona" (Clarke 1994), "dividuals" (Deleuze 1992), or "data doubles" (Lyon 2003) are used to describe the comprehensive compilation of personal data, the creation of increasingly detailed and fine-grained digital footprints of individuals, which are then later processed in machine learning applications, which in turn have various (and in some cases negative) ramifications for society (Calvo et al. 2020; Eubanks 2018; O'Neil 2016). In this process, various "filters" mediate the translation from an individual's original behavior to eventual computer outputs. Those filters take effect through the selection of certain sensors, data cleansing processes, feature extraction, software libraries, data visualizations, etc. This is why the literature on critical data studies claims that something like "raw data" does not exist (Gitelman 2013). There are, metaphorically speaking, always bottlenecks, strainers, gates, intentionally or non-intentionally regulating the "permeability" for data at different stages of the computational processing of reality. But those filters do not have an ethical dimension. They do not lead to an ethically motivated selection and sorting out of different data contexts with varying ethical data qualities. To define what I mean by ethical data qualities, one has to analyze how data quality is affected by certain personality traits or modes of behavior of individuals, and how those traits or states can be assessed from an ethical point of view. Eventually, finding quality data shall not serve the pursuit of an improved marketability, but of socially accepted, beneficial machine learning applications.

# 3 Human behavior and its digital records
## 3.1 Classifying human behavior

Typically, behavioral data are the result of tracking online activities of all kinds. Different modes of behavior eventuate in different data contexts. Individuals leave different data traces behind depending on their emotional state, educational background, intelligence, wealth, age, moral maturity, and the like. In order to sort those traits and to classify human behavior and stages of development, one can draw on well-established theories in psychology and sociology. Within the framework of these theories, the aim is to distinguish different modes of behavior or stages of development according to empirical findings. As a general rule, behavior or personality development is understood to be largely a product of one's social environments. Those environments are classified, for instance, with the help of theories of social stratification (Grusky 2019; Vester 2001; Schulze 1996; Erikson et al. 1979; Bourdieu 1984). A person's milieu, meaning, simplistically speaking, upper, middle, or lower classes, determines their habitus, which in turn determines large parts of their behavioral routines, and vice versa. Individuals occupy a certain position in "social space" which is the result of a contested distribution of resources, meaning economic, cultural, social, or symbolic capital (Bourdieu 1989). The position an individual occupies in social space is in large parts "hereditary" and can be affected by social injustices. Nevertheless, the amount of capital a person can concentrate on her- or himself has a structuring power on many areas of life, meaning that it organizes a person's taste, language, estate, political orientation, or, to say it more generally, his or her dispositions.

Further, these dispositions also structure and determine the way a person uses digital technologies, and influence what kind of data are tracked by these technologies. By using terms like "media-based inequalities", "digital divide" or "digital inequality", several studies show the strong influence a user's socioeconomic status has on media or Internet usage patterns (McCloud et al. 2016; Zillien and Hargittai 2009; boyd 2012; Hargittai 2008; Mossberger et al. 2003). Individuals with a higher socioeconomic status are more likely to engage in online activities that enhance their social position, have status-specific interests, interact more frequently with e.g. political or economic news or health information, have higher levels of computer literacy, use less often chat platforms or social networking sites, and so forth. All in all, the position of an individual in social space heavily influences his or her ways of using digital technologies and hence the kind of behavioral data that are digitally recorded – with the respective consequences for biases, scopes, representative statuses, or ethical quality dimensions of data sets.

While behavior is in many respects an outcome of the respective social environment, class, milieu, or social position, the same holds true for personality development, which is widely dependent on the circumstances of socialization. Developmental psychological theories postulate that personality development passes through various stages, where logical reasoning is learned, moral senses are developed, social norms are adopted, emotional intelligence is acquired, stereotypes are negotiated, role models are changed, self-reflection is learned, values are internalized, personal crises are overcome, and the like (Erikson 1980; Kohlberg et al. 1983; Loevinger 1997). In order for an individual's socialization to succeed, it requires, among other things, a certain range of beneficial influences from a social environment, which can be separated from harmful influences. To scrutinize these influences is the objective of developmental psychology. The discipline focuses on long-term progressions with regard to the experiences and the behavior of individuals in order to find patterns and regularities that are crucial for the development of intellectually and emotionally sound and mature individuals (Lener 2015). A succeeding development is measured by aspects such as problem-solving abilities, emotional intelligence, cognitive development, prosocial behavior, mental health, educational success, etc. As soon as such norms for a successful personal development are defined, one can roughly differentiate between positive and negative environmental influences. The latter can affect health, gross and fine motor skills, socio-emotional development, the speed of information processing, self-concepts, knowledge, or language behavior and range from alcohol to stress during pregnancy, residential areas with high crime rates, low educational levels, emotional, physical or sexual abuse, as well as a neglectful parenting style (Sullivan and Knutson 2000; Spera 2005).

Crucial is the fact that developmental stages correlate with actual behavior. "Higher" forms of personality development lead to other behavior patterns than "lower" ones (Hart et al. 1997; Paul B. Baltes et al. 1978; Kohlberg et al. 1983). Normally, more cognitive-moral growth leads to more socially desirable or acceptable behavior. Philosophical theories about ideal moral acting, ranging from Kant's categorical imperative (Kant 1977), Habermas' discursive will-formation (Habermas 1987), or Rawls' theory of social contract (Rawls 1999), imply that individuals possess fully developed cognitive capacities. Apart from the fact that these philosophical models have no anchoring in the empirical reality of the social world or the human psyche (Willke 2005), one can assume that personality or character development may strive towards the target values and rationality standards of these models. In order to measure the "proximity" of a person's character to certain target values, differential psychologists have developed various tools for personality assessment to understand and predict behavior in different social contexts. Amongst personality assessment tools are the widely used Five-Factor Model (John et al. 2008; McCrae and John 1992), non-scientific tests like the Myers-Briggs type indicator (Myers and Myers 1995), or less known methods like the Multidimensional Personality Questionnaire (Tellegen and Waller 2008), "PerformanSe" (Patel 2006), and many more. All those tools have specific weaknesses, they ignore the fact that personality can be in a state of flux, and that it may be unclear what personality characteristics mean in terms of behavioral manifestations in certain situations. But apart from that, they more or less reliably measure traits like motivation, extraversion, emotional stability, openness, conformism, rationality, impulsivity, dynamism, anxieties, social activity, and the like.

The mentioned theories and tools have a tacit consensus about certain ethical target values or "attraction poles" (Sloterdijk 2009). This can be exemplarily elucidated with regard to the Five-Factor Model (John et al. 2008). All five personality dimensions have an attraction pole, meaning that all dimensions can be spanned between two poles whereas one pole is designated as the favored one. Typically, more complexity in an individual's mental and experiential life is better than less (openness), more impulse control that facilitates goal-directed behavior is better than less (conscientiousness), more social activity and positive emotionality is better than less (extraversion), a more prosocial and communal orientation is better than less (agreeableness), and more emotional stability is better than less (neuroticism). In the background of all personality assessment tools, developmental psychological, or social milieu theories, tacit normative presuppositions exist that structure attraction poles of all kinds. However, making these presuppositions and polarizations explicit may not be equated with attempts to classify humans as such. The mere idea of classifying humans provokes strong moral intuitions to refuse such practices. But besides these moral intuitions, the application of classification or scoring systems of all kinds on humans is common industry and government practice in many countries (Engelmann et al. 2019). People are classified with respect to their financial situation, their social reputation, their risks of conducting certain actions, their personality, etc. Moreover, they are classified along geodemographic segmentations, purchasing histories, lifestyle types, and the like. However, the circumstance that the application of ranking systems on people corresponds to the status quo of the digital economy does not mean that the related practices are morally correct. Quite the opposite may be true (Zuboff 2015). Nevertheless, in order to frame digital ranking practices in a more politically correct manner, one should stop saying that individuals as such are classified but that particular types of behavior or particular personality traits are ranked with respect to certain dimensions. The fact that all these dimensions have a more or less strong normative alignment shows that individual- or behavioral-related ranking methods are always morally relevant, but at the same time they are an essential part of human interaction and organization. The crucial questions are, though, how one can measure and classify personality traits and different types of human behavior via computers. And how can these practices be used for beneficial purposes?

## 3.2 Tracking human-computer interaction

Dataveillance (van Dijck 2014; Clarke 1988), in other words practices of recording and analyzing digitally mediated behavior, has at least three complications or downsides. First, it is a morally contested practice, causing negative "chilling effects" of all kinds (Schneier 2015). Second, monitoring human-computer interactions or online behavior does not yield data that corresponds to real attributes but it constructs them (Haggerty and Ericson 2000). And third, one can only infer personal attributes with the right data bases. Obviously, missing data strictly limits the scope of information one can gain. One has to differentiate between deliberately displayed expressions that can be digitally recorded, and statistically inferred information from those records. All behavioral detection techniques are never perfect but approximately right. Especially information that is not actually recorded but statistically predicted from data sets is always probabilistic, albeit in many cases very accurate. Before these three complications or downsides are discussed in more detail in chapter 5, it is important to see that in practice, various tools are used to track various personal traits and states (Matz and Netzer 2017). Single- as well as multimodal approaches that combine several psychological attribute recognition methods, and that can detect involuntary (e.g. physiological), semi-voluntary (e.g. facial expressions), as well as voluntary (e.g. key presses) signals are used (Zeng et al. 2009; D'Mello and Kory 2015).

  Some instances are listed hereafter: By analyzing clickstreams, browsing histories, or search queries, inferences on users' demographic information can be made (Acar et al. 2014; Hu et al. 2007; Bi et al. 2013). Affective computing serves to detect emotions, mostly through text, voice, face, or posture processing techniques (Picard 1997). Written text can be investigated in order to detect mental illnesses, to conduct sentiment or personality analysis via differential language analysis, natural language processing, and other machine learning methods (Guntuku et al. 2017; Schwartz et al. 2013; Pang and Lee 2008). Moreover, digital images, for instance social media profile pictures, can also be used to reveal personality attributes (Segalin et al. 2017). Various sensors – especially the ones in smartphones and other wearable devices – are used to track physiological signals, movements, activity levels, mobility patterns, face-to-face encounters, and the like in order to infer internal states and personal attributes (Harari et al. 2017; Kwapisz et al. 2011). User input via display touching behavior, mouse movements, or keyboard strokes can also be used to infer personality traits (Khan et al. 2008). Many other applications could be added. For a cursory overview of types of data traces and the various inferences that can be drawn from them, see figure 1.

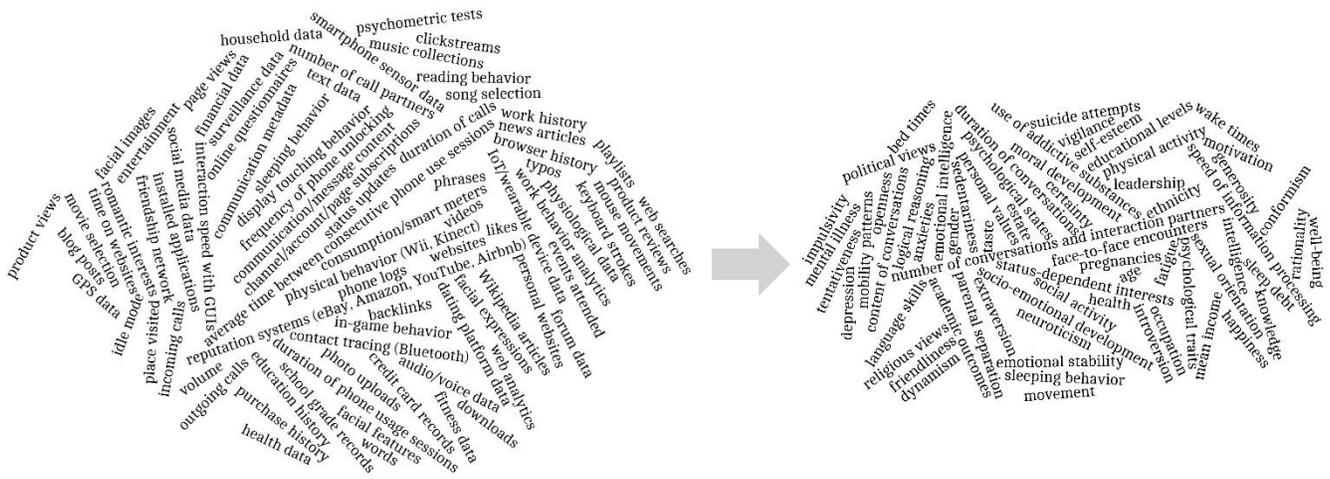

*Figure 1 - An exemplary overview of types of data traces and the various inferences that can be drawn from them*

  All in all, the baseline is clear. Tracking technologies for digitally mediated behavior can in many cases successfully measure a broad spectrum of psychological traits, affective states, and personal attributes. Many tracking applications specifically aim at measuring the six basic emotions (sadness, fear, anger, disgust, joy, surprise), while in practice, though, those basic emotions can be observed only seldom, instead engagement, confusion, boredom, curiosity, frustration, and happiness are more frequent affective states in human-computer-interaction (D'Mello 2013). But tracking technologies can also measure more complex attributes like age, gender, sexual orientation, occupation, mean income, ethnicity, religious views, political attitudes, personality traits, intelligence, pregnancies, use of addictive substances, job performance, parental separation, and many more (Kosinski et al. 2013; Kosinski et al. 2014). Here, the connection to psychological as well as sociological approaches to classify human behavior can be made. While many classical theories from the humanities approach social structures and individual-related traits and distinctions on a very high and abstract level, digital behavior tracking technologies only capture "microscopic" behavior traces in data sets. But the former and the latter can be combined in order to transition from the *n = all* ideology to singling out datasets from certain subpopulations that are deemed to be the most competent or morally versed group for a particular task.

Tracking technologies, even if they are themselves ethically contested, can be the bedrock for an ethically motivated selection of different data contexts with varying ethical data qualities that can then lead to beneficial machine learning applications and machine behavior. When recognizing that a person's dispositions structure and determine the way she or he uses digital technologies, then methods to detect digitally mediated behavior can, in turn, infer those dispositions when analyzing data traces. That means that "higher" states of personality or moral development, socially desirable or acceptable behavior, distinct cognitive abilities, emotional stability, rationality standards or, in general, the "proximity" of a person's dispositions to certain socially accepted and ethically defined target values can be measured. Problem-solving abilities, emotional intelligence, cognitive development, prosocial behavior, educational status, mental health – all those assessment dimensions have attraction poles that are used in many social contexts to rank human behavior and to assess whether particular individuals can be put in charge, or are competent enough or eligible for certain tasks. This principle is, at least when adopting a meritocratic perspective (Young 1994), effective in many social institutions. From here on, I want to make the transition to prefiltering training stimuli for machine learning applications according to certain individual attributes, states, and traits.

## 4   Beneficial machine learning: putting the approach to practice

In the context of (supervised) machine learning development, there are three ways in which hereditary and "environmental" information can be inscribed into algorithms: they can be incorporated into applications by programmers making design choices in algorithms (Brey 2010; Friedman and Nissenbaum 1996), by particular training stimuli (i.e. data), or by a machine's own "experiences". Taking up the humanistic differentiation between hereditary factors and environmental influences that shape an individual, one can stress that machine learning applications also combine both; the former through algorithm design, and the latter through training stimuli, where both factors interfere with each other. Training stimuli, in other words a set of examples used for learning, are used to fit and tune the architecture, parameters, or weights of a classifier. Training data sets are supposed to allow artificial neural networks to generalize from the sample of the training data set to potentially every other case, meaning that the network has the best possible performance on any new data. In this context, training stimuli must be distinguished from validation and test or holdout sets, where the former serve the purpose of tuning the architecture of a classifier, and the latter of measuring the performance of a trained classifier. Supervised machine learning predicts a categorical or continuous value $Y$ in the form of target variables or labels given an input $X$ in the form of a set of variables by training a function $F$, where $F(X) = Y$ (Ghani and Schierholz 2017). Here, the set of variables can represent behavioral data, but one also has to keep in mind that the same holds true for labels. Labels are most often the result of manual clickwork (Irani 2016), but they can have the form of genuine behavioral data, too. For instance, this is the case when labels for video data of an autonomous vehicle's surroundings are generated by capturing the driver's behavior (Tsutsui et al. 2018). This way, human behavior also becomes machine behavior via labels, not just via training data itself.

Concepts that follow the idea of an ethically motivated selection or limitation of particular training data or labels in order to influence the process of developing certain machine behaviors are non-existent up to this point. One single exception is Davidow and Malone's cursory concept of "starving AI" or of "putting artificial intelligence on a data diet" (Davidow and Malone 2020). The idea is to ensure trustworthy artificial intelligence not by controlling it, but by putting it in "virtual prisons", meaning that the applications are disallowed to use whatever training stimuli they can get for learning. The authors follow a rather metaphorical approach, but, in a nutshell, they rudimentarily capture an idea similar to the one presented in this paper, namely developing beneficial machine learning applications by filtering training stimuli according to ethical considerations. My line of argument starts at the assumption that a person's social background, educational level, personality, intelligence, etc. shape his or her way of using digital devices. Moreover, these devices are equipped with sophisticated tracking technologies that can in many cases accurately measure and infer the user's personal attributes, traits, or states. Depending on these measurements, data traces the respective user produces, that is data traces that provided the basis to the measurements itself as well as data traces that are situated in the same context as the measurement, are assessed from an ethical perspective. Thereafter, this assessment enables data scientists to relinquish the idea of using as much relevant data as possible that represent averages of whole populations. Instead, they single out quality data that are representing behavior of specific subpopulations which are deemed to be especially competent, eligible, or morally versed for a particular task. This limitation, that stands in contrast to the credo that the bigger the data the better the machine learning models, serves to tailor training data in a way that machine behavior can be steered into a direction that promotes its beneficence.

## 4.1 Concrete use cases

To further elaborate on that, I want to sketch out use cases that exemplarily illustrate the process that is outlined above. For that end, I investigated five machine learning driven applications and demonstrate how beneficial machine behavior objectives can be achieved by selecting certain quality data contexts for model training. The investigation shall delineate how the paper's ideas can be put from theory to practice. I will focus only on applications that are widely used, like e-commerce recommendation systems, search engine ranking algorithms, or autopilots in self-driving cars, and describe how these applications can be amended by following a stringent approach for quality training data selection along particular ethical considerations.

### 4.1.1 Self-driving cars

#### 4.1.1.1 Machine behavior objectives

Autonomous vehicles are supposed to guarantee as much safety as possible (Koopman and Wagner 2017). Avoiding crashes with self-driving cars (Xu et al. 2019) is paramount to advance their deployment. Achieving this goal has many dimensions, but it certainly encompasses safe machine behavior, meaning a car complies with safe overtaking maneuver, following, emergency stop, cornering, or line choice rules.

#### 4.1.1.2 Data sources

Autonomous cars must infer from past traffic situations to new ones. Thus, training data, meaning video recordings and further sensor data of all kind representing countless hours of driving, as well as annotations for these data are of utmost importance. Because it is rather expensive and in some cases notoriously difficult to acquire enough annotation data, in some autonomous vehicles, label collection happens via measuring behavioral cues from human driving behavior, e.g. acceleration, deceleration, steering, etc., in manual mode or during autopilot disagreement (Eady 2019). The labels are then linked to the respective footage of the vehicle's surroundings. Additionally, data traces from actual driving can be combined with customer data as well as further behavioral data from third party organizations like data-brokers.

#### 4.1.1.3 Tracked states and traits

There are certain individual characteristics like gender, age, driving experience, distraction, attention, reaction time, visual function, sensation seeking, impulsivity, etc. that predict risky driving behavior (Fergusson et al. 2003; Wayne and Miller 2018; Anstey et al. 2005). According to accident statistics and empirical investigations, individuals who cause fatal as well as non-fatal car crashes tend to be male, of young age, have high levels of aggressiveness, sensation seeking, and impulsivity as well as some other traits like lower levels of income, poor mental health status, higher levels of neuroticism, possibly raised blood alcohol concentration, lower driving experience, and show various forms of antisocial behavior or higher levels of social deviance (Čubranić-Dobrodolac et al. 2017; Vaughn et al. 2011; Abdoli et al. 2015; West and Hall 1997; Hyman 1968; Wang et al. 2019). Many, if not all of these traits can be digitally detected at some degree of accuracy. Those characteristics as well as additional cues like engine speed, pedal pressure, improper following, speaker volume, driver body posture, gestures, head movement, verbal outbursts, etc. can be digitally tracked in order to predict a driver's safety level (van Ly et al. 2013).

#### 4.1.1.4 Quality data contexts

As soon as the above-mentioned traits are digitally tracked and recorded, the driving behavior data that is related to the respective driver can be excluded or downgraded from the data set that is used to train the models that determine the machine behavior during autopilot. Traffic psychologists can help machine learning practitioners to further establish tools to classify data that represent decent driving behavior. In short, quality data contexts arise from drivers who possess decent driving experience, have a good reaction time, tend to be female, have low levels of aggressiveness, sensation seeking and impulsivity, show active head movement in traffic, distinguish oneself in few or no verbal outbursts, proper following behavior, or modest acceleration behavior, to name just a few attributes.

### 4.1.2 Language generation
#### 4.1.2.1 Machine behavior objectives
Chatbots as well as speech assistants of all kinds are supposed to produce appropriate, sufficiently eloquent language that does not violate social norms, discriminate against certain groups of people or perpetuate biases that are incorporated into training data (all of which is especially precarious in open domain conversations) (West et al. 2019; Danaher 2018; Silvervarg et al. 2012; Sheng et al. 2019; Bolukbasi et al. 2016).

#### 4.1.2.2 Data sources
Natural language generation is based on finding statistical patterns in text corpuses (Solaiman et al. 2019), which then allows a machine learning model, among other things, to predict the next word in a sentence based on previous words. To learn those patterns, the chosen text corpuses can be digitized books, forum posts, news articles, communication data, Wikipedia articles, websites, blogs, scientific papers, and many more.

#### 4.1.2.3 Tracked states and traits
States and traits that can be tracked in order to assess text data quality may range from an author's educational background or occupation, intelligence, the characteristics of his or her keyboard strokes or display touching behavior (backspacing etc.), the time between writing and posting, and in particular by assessing the used publication platform, filtering intermediates, review processes, and the language skills themselves.

#### 4.1.2.4 Quality data contexts
Especially text data that is not produced by professionals, meaning journalists, writers, scientists, etc., but by lay persons is expected to be of lower quality. Text data that is not editorially controlled and therefore did not undergo any kind of review or filtering intermediate may be interspersed with orthographic mistakes, poor syntax, smaller word pools, slang, invectives, strong biases, etc. Quality data contexts are to be assessed in dependence on the respective purpose of an application for natural language generation. Texts from the public domain may be suited to improve a chatbot's realism, hence its ability to produce convincing, authentic, and human-like everyday language. On the other hand, these texts can be infiltrated with aggressive, discriminatory, or offensive phrases. To avoid these and other pitfalls, the selection of text corpuses that are used to train conversational robots should not follow the bigger-is-better-approach like many commercially developed chatbots do. Instead, the selection of corpuses can be narrowed down to digital writings that underwent a firm quality check through publishers, peer reviews, or media agencies, that is embedded in a sophisticated web of citations or links, or that stem from individuals with high levels of language skills. Moreover, language proficiency can be determined by assessing the structure, continuity, errors, vocabulary richness, length of sentence, changes made to text, etc.

### 4.1.3 Search engines
#### 4.1.3.1 Machine behavior objectives
Modern search engines like Google Search, Bing, Yandex, etc. use a plethora of signals to rank search results, make autocomplete suggestions, predict users' intentions, evaluate websites, and so on. The main machine behavior objective is to ensure that rankings and content fit to the anticipated needs of the users. This, in turn, is supposed to cause a lock-in-effect and bind users to the respective search engine, eventually raising the likelihood of contact with advertisements. Despite this well-established machine behavior objective, one can name several other objectives that could determine the architecture of search algorithms like content quality, expertise, and trustworthiness (in general search engines), equal opportunity (in people search engines), sustainability (in product search engines), and many more.

#### 4.1.3.2 Data sources

Search engines use diverse tracking techniques, harnessing the large amount of different human-computer-interactions. Each list of search results shown to users nudges them to become behavioral data contributors for further calibration and model training by clicking on links, mousing over items, using the back button, scrolling through pages, entering terms in search bars, interacting with ads, spending time on a page, and many more. Besides such behavioral data, search engines can analyze main and supplementary contents of websites, amount of internal and backlinks to a website, labels, page load speed, aggregated views, end-user device specifications, duplicates, and so on.

#### 4.1.3.3 Tracked states and traits

Many of today's relevant search engines are embedded in broader online platforms that allow for a comprehensive user classification. By collecting and analyzing data on search terms, visited websites, clicked ads, user location, keyboard strokes, mouse movements, interaction speed, product or profile views, and the like, it becomes possible to probabilistically infer a bunch of different personal states and traits. Among them are a person's gender, age, occupation, residence, religion, political views, favorite brands, personality, intelligence, literacy, and many more (Bi et al. 2013). These states and traits can then be used to assess a user's "signal quality".

#### 4.1.3.4 Quality data contexts

Professional general search engines do have page quality rating systems in place (Underwood 2015). They are used to recognize the purpose of a website. Beneficial pages that are supposed to help users and are created by individuals with high expertise, authoritativeness, and trustworthiness receive the highest ranking. Pages that contain hate or misinformation, encourage harm to others, or have a deceptive intent receive the lowest rating. However, these page quality ratings do not solely determine the search results. They are accompanied by machine learning techniques that "digest" user behavior in order to re-train the search algorithm. This user behavior can also be assigned to varying "quality" stages. The clickstream habits of a person who, for instance, regularly uses politically extremist search terms, visits websites of low quality, has numerous typos, etc. should be less considered for shaping the search algorithm. On the other hand, clickstream habits that give evidence of ethically desired traits could preferably be used to optimize ranking algorithms.

### 4.1.4 Social media

#### 4.1.4.1 Machine behavior objectives

Recommendation systems on social media platforms come in all shapes and sizes. They are used to filter posts, friends, images, videos, music, news, search results, and many more. Hitherto, the main goal of these systems is to increase user engagement in order to bind them to the respective platform. This, in turn, shall raise the likelihood of advertisement contact and click-through-rates (Kuss and Griffiths 2017; Eyal and Hoover 2014; Hagendorff 2019b). Taking social responsibility seriously, platforms could rearrange their objectives towards values of a vital and fair public discourse, truth, and information quality. This means to change the methods for algorithmic measurement and determination of information relevance. Fake news, hate speech, extremist content, etc. may cause the strongest user engagement, but the engagement quantity should not determine the subsequent dissemination and recommendation of the respective content. Instead, engagement quality should determine data quality and help to build responsible machine recommendation behavior.

#### 4.1.4.2 Data sources

Social media platforms can track a plethora of user signals. Amongst the more obvious ones are clickstreams, search queries, demographic or profile information, reactions to posts, duration of post views, scroll behavior, networks of friends, comments, and many more. All these data traces are used to determine the relevance of posts, videos, images, tweets, friend suggestions, etc. in order to operate the platforms' recommendation systems.

#### 4.1.4.3 Tracked states and traits

Tracing back to a differentiation from behavioral economics (Kahneman 2012), one can distinguish system-1- and system-2-interactions. System-1 comprises fast, emotional, effortless, cognitively simple thinking processes that are prone to biases and mistakes, whereas system-2 covers slow, rational, and deliberate thinking processes. Those two modes of thinking do also influence the way digital platforms are used (Lischka and Stöcker 2017). Amongst other factors, social media platforms could measure whether users operate with a platform on a more irrational, bias prone, impulsive system-1 mode. This mode allows for rather quick and impulsive actions, resulting in a stream of unreflected human-computer-interactions. Impulsive, system-1 user behavior could be tracked by things like reaction or comment speed, the susceptibility to nudging techniques, and scrolling or reading behavior. The platforms could also use further inferences to educational levels, intelligence, psychological traits and states like anxieties or radical political or religious views to assess user behavior that is connected to the respective attributes.

#### 4.1.4.4 Quality data contexts

Cognitive heuristics that are part of system-1-interactions influence the way individuals interact with social media. Hence, biases are technically perpetuated via recommendation systems (Stieglitz and Dang-Xuan 2012, 2013). Quality data can thus be scraped from contexts where user generated data do mainly represent system-2-human-computer-interactions. This way, recommendation systems can be trained on behavioral data that represents fewer biases and impulsive reactions. Instead of negatively affecting public discourse by helping the spreading of content that is mostly suited to cause emotional arousal and impulsive reactions, platforms help to automatically disseminate content that is less "toxic" for public discourse. In this context, a special focus can be laid on so-called "superusers", who are not just very active users with high levels of engagement, but who also disproportionately spread misinformation. According to one rare source, internal committees at Facebook urged to lower recommendation scores for content posted by "superusers" on the far right or far left of the political spectrum. Content from moderate users, in turn, would receive higher scores (Ng 2020). This advice was turned down by Facebook's leadership. However, it perfectly corresponds to the idea that is advocated here. When recommendation systems are trained on behavioral data from individuals with higher education, who do not represent political or religious extremes, and who normally interact with quality, i.e. journalistically or scientifically verified, trustworthy information, it is to be expected that those systems automatically spread content that comes with various benefits for the public, instead of harms, leading to a situation where everyone is better off.

### 4.1.5 Online shopping

#### 4.1.5.1 Machine behavior objectives

E-commerce platforms where people can buy goods and services use various methods to promote purchasing behavior. They use shopping search engines, product recommendation systems, product reviews, dynamic pricing, cross selling, customer analytics tools, conversion rate optimization, conversion funnels, varying payment options, specific user interface designs, etc. in order to maximize revenues. Therein lies the entrenched main machine behavior objective. However, via tweaking the underlying machine learning algorithms, the machine behavior objective can be diversified, comprising not just the pursuit of economic values, but also values of sustainability or public health. Especially the biggest online stores like Amazon, eBay, Walmart, Jingdong, Alibaba, and others would cause a significant impact by just slightly changing their machine learning models towards the mentioned values, taking their corporate social responsibility seriously.

#### 4.1.5.2 Data sources

No different than search engines or social media platforms, online retailers can collect a broad variety of user data. They can analyze and track the number of transactions, all kinds of product and customer data, the conversion rate, product impressions, average order values, product detail views, adding or removing of products from the shopping basket, withdrawals from checkout process, customer lifetime values, traffic sources, details of users' devices, and many more.

#### 4.1.5.3 Tracked states and traits

Using the many data sources e-commerce platforms can gather as a basis, they can implement specific automated mechanisms for customer segmentation. Typically, differentiations for types of customers are made purely from a sales perspective, distinguishing between loyal, impulsive, novice, etc. customers. Notwithstanding that, customers can be segmented along criteria like health- or eco-consciousness by analyzing their product views, shopping behavior, product reviews, search terms, personality, socio-demographic factors, and the like.

#### 4.1.5.4 Quality data contexts

Sticking to the aforementioned values of sustainability and public health, e-commerce platforms could use all available data from health- as well as eco-conscious customers and use specifically those data to train models for product recommendations, dynamic pricing, or ranking algorithms for their search engines, to name just three major setting options. Such measures could significantly foster the extent to which e-commerce platforms promote more sustainable and healthier consumer behavior.

# 5 Discussion

In the following section, I want to gather some major points for discussion to the suggested approach to achieve beneficial machine learning applications. The approach relies on several practices like tracking, profiling, ranking, or filtering that have been applied in contexts of technology misuse for illegitimate ends more than just a few times (Brundage et al. 2018; Crawford et al. 2019). Nevertheless, the mentioned methods cannot be simply discarded. Rather, they have to be used for purposes that are in line with the common good and ethical values. The following subchapters will elaborate on that and take up the seven most pressing concerns that can be raised when putting the paper's ideas into practice.

## 5.1 Ranking behavior

First of all, I want to address some ethical concerns that are connected with the idea of ranking human behavior, as described in chapter 3. As a cautionary tale how such ranking practices fail one could point at digital marketing, where marketers consider customers to be either "targets" or "waste" (Turow 2012). People's tastes, demographic profiles, beliefs, income levels, and many more are digitally measured for the only purpose of discriminating them for commercial goals and to find the customers that are deemed most valuable. Reputation silos are constructed around people who statistically seem similar, but this practice often circumvents ethoses around equal opportunity, justice, or transparency. However, the idea of ranking or sorting human behavior along certain dimensions or hierarchies is not to be refused altogether. One can put the whole concept in another light when reframing it according to assessment criteria from ethics. Hence, besides just arguing for a repurposing of existing social sorting structures in marketing for other socially accepted ends, one can in general say that society embraces various practices of behavior assessment under the term "ethics". Here, one transitions seamlessly from saying that some social positions, dispositions, or stages of development are better than others to classifying behavior as morally right or wrong. In the context of this paper, though, the argument is somewhat more specific: Currently, digital tracking systems that measure users' social as well as psychological traits and states are used mainly to support marketing decisions, to foster customer relationship management, to personalize the marketing mix to individuals, and to support many other commercial purposes (Wedel and Kannan 2016). I argue that the rich toolset that is already established for marketing purposes can be repurposed to assess the ethical quality of data contexts in order to develop beneficial machine learning applications. Currently, tracking methods are not deployed to evaluate digital behavior from an ethical point of view, which would be essential to assess ethical data quality dimensions. With that said, this assessment does not require detailed knowledge about ethical theories that are developed in philosophic discourses. Apart from a few complicated, dilemmatic cases, moral intuitions can appropriately guide ethical reasoning, or, to be more precise, ethical judgements are driven primarily by one's intuitions (Haidt 2001). This becomes evident when considering the tacit consensus about ethical target values or "attraction poles" that are embedded into sociological as well as psychological theories of all kind. Normative assumptions about the value of prosocial dispositions, rationality, moral development, openness, impulse control, positive emotionality, and the like are hardly contested. Hence, these normative assumptions can guide data quality evaluations. Hereafter, particular data sets can be used as training data for beneficial machine learning. All this does not happen in order to discriminate against certain groups of individuals, or to sort people for the purpose of improving businesses. Rather, it happens for the purpose of developing beneficial machine behavior in order to foster the common good.

## 5.2 Paternalism

Another potential objection to the ideas presented here is that they depict a form of ethical paternalism. The decision about what defines quality data contexts is made by machine learning practitioners, and it affects other collectives without their democratic consent. The idea of paternalistic thinking is that some individuals are more competent, rational, or versed than others and that the former can decide for the latter for their advantage. Paternalistic approaches are mainly criticized because they not only limit the freedom of affected individuals, but this is also done without their consent (Sartorius 1983). In this context, two main counterarguments can be raised. First, the idea of developing beneficial supervised machine learning applications follows, as the term suggests, the notion of being beneficial or advantageous for as many individuals as possible, hence promoting the common good. Second, machine learning applications do in nearly all cases affect individuals without their explicit consent or knowledge. Values are part of technologies itself (Brey 2010), and the process of embedding certain values or choosing architectural designs is in many regards not a democratic one. Rather, a small group of technology developers possess the power to make far reaching decisions for a group of end users that can comprise millions or billions of individuals (Lessig 2006). The decisive question is whether values are embedded in software in an arbitrary and nonreflective way, often perpetuating prejudice, biases, or misunderstandings, or whether these values and ethical norms are chosen carefully and consciously. Here, I opt for the latter. In addition to that, technology ethics indeed typically operates with paternalistic, top-down norms and standards. Asimov's Three Laws for Robots, for instance, is presumably the most well-known approach (Asimov 2004). However, in the context of this paper, the paternalistic top-down approach is not solely embraced but combined with one that is bottom-up. Bottom-up approaches mean that a technical agent, in this case machine learning models, explores courses of action that represent morally praiseworthy examples (Wallach and Allen 2009). Hence, the agent achieves "moral capabilities" by surveying its environment, similar to childhood development. Here, both top-down and bottom-up approaches are integrated. A top-down analysis is done with regard to the way training data is filtered or selected according to ethical criteria, while on the other hand a developmental or bottom-up approach is chosen in order to allow the machine to learn a certain behavior from data sets.

## 5.3 Transparency

Another objection, that is akin to the one above, is to stress that the proposed approach for beneficial machine learning applications is non-transparent, leaving affected individuals unwitting about the technical measures that are conditioning their user experience and filtering mechanisms for quality data. The counterargument to this objection is that platforms or software developers could and should have no problems whatsoever making transparency statements, thereby informing potential reviewers about the value and design choices, ultimately making them subject to public scrutiny. This would show that definitions about ethical quality dimensions of training data are in line with cultural consensuses, ethical theories, and moral intuitions. Revisiting the above-mentioned examples from chapter 4.1, the majority of people would consent to the claim that autonomous cars should be safe, that natural language generation should avoid simplicity and perform eloquently, that e-commerce platforms should promote sustainable shopping routines, that recommendation systems on social media platforms as well as search engines should not foster political extremism and information that is "toxic" to the public discourse but promote quality content, expertise, and truth. In general, it should be a necessary prerequisite that through transparency statements the criteria by which data quality is evaluated are described and justified when putting technologies for digital behavior tracking as well as machine learning models that are trained with behavioral data in place.

## 5.4 Privacy

In order to select for quality data or data that represents ethical behavior, a bunch of tracking or surveillance techniques must be in place. This leads to a further objection, namely that the proposed approach for beneficial machine learning application goes hand in hand with privacy violations. This is true and cannot be disagreed with. At least, the proposed approach does not necessarily opt for an extension of methods for recording behavioral data that are already entrenched (D'Mello and Kory 2015; Zeng et al. 2009). This is a weak excuse, but nevertheless, one can stress that decisive is that these methods are used for legitimate ends, not that they are abolished (Belliger and Krieger 2018; Hagendorff 2019a). As already discussed in a previous chapter, the idea of classifying people or people's behavior raises weighty ethical questions and has its ailments, especially with regard to the feasibility of data-driven assessments of sensitive traits like mental illnesses, intelligence, personality, and the like. However, data protecting measures that would prohibit these assessments like the one mentioned are primarily aiming at preventing unjust discrimination and at securing personal autonomy (Roßnagel 2007). In the end, though, is it important to remember that from the pure existence of these technologies alone it does not necessarily follow that they are misused. On the contrary, when binding legal norms as well as strong ethical tenets are entrenched, tracking and profiling can be used for the common good, as it is described here. Moreover, techniques for tracking user behavior and assessing behavioral data quality can work by only using anonymized and aggregated data, avoiding any opportunity to identify certain individuals contributing to or being excluded from contributing to training data sets.

## 5.5 Accuracy

Another objection is that techniques to digitally assess and rate human behavior may be inaccurate and create false positives and negatives. These techniques as well as data traces per se construct rather than represent an individual's true actions, traits, and states. Accordingly, behavioral data as well as the computational processing thereof cannot be condensed in information that represents "reality". Rather, different "realities" can be constructed from data and algorithms (Matzner 2016; Lewis 2015). They do not work impassive, but shape how we understand the world in a performative manner (Kitchin 2017), while allowing probabilistic inferences on in situ behavior. In individual cases, this can lead to detrimental false positives or negatives. But the methods proposed in this paper all operate with aggregated and, in the best case, anonymized data, which means that individuals face unjust technological consequences only when tracking and profiling techniques fail significantly. Only in the unlikely event they come up with misclassifications in an overwhelming number of cases, the selection of quality training data and therefore the trained models would become skewed.

## 5.6 Discrimination

Akin to the aforementioned objection is the argument against algorithmic discrimination. The ideas presented in this paper could fall prey to such arguments since specific biases in data sets are intendedly promoted, resulting in "skewed" algorithmic decision making. Previous discourses on algorithmic discrimination rightly criticize that machine learning techniques perpetuate existing biases that are entrenched in data sets and therefore foster unfair discrimination. Under the umbrella term "fairness, accountability, and transparency in machine learning" (FAT ML), machine learning practitioners collect methods for reducing algorithmic discrimination primarily by dealing with protected attributes like gender, age, ethnicity, etc. (Dwork et al. 2011; Kleinberg et al. 2016; Veale and Binns 2017). However, here I argue that one should reintroduce or promote "algorithmic discrimination", but, needless to say, not in the traditional way. Data sets may contain features that are critical in a way that they should be weighted stronger than others. This can be the case even when these features perpetuate bias. Biases can be necessary for fairness and the proper functionality of certain processes. Dutta et al. (2020) give the example of a hiring for fire fighters, where candidates should be able to lift heave weights, which leads to a preference for men rather than women. In short, biases are acceptable if they are critical for the legitimate solution of a given task. Here, I propose to promote biases in data sets used to train machine learning models that lead to a preferability of features that are desirable from an ethical point of view. At the same time, however, this means to put individuals at a "disadvantage" who produce behavioral data that originates in deeds, personal traits, or mental states that are socially less esteemed like risky behavior, detrimental norm violations, bad language, low education, political or religious extremism, flawed logical reasoning, impulsiveness, and the like. This way, machine behavior that results from recognizing statistical patterns in behavioral data is not "socialized" by general populations ($n = all$) but by specific subgroups that comprise individuals who are most competent, eligible, or morally versed for a particular task. While the typical notion of algorithmic discrimination is pointing towards unfair computational outputs, the kind of algorithmic discrimination that is proposed here aims at introducing stricter filters that thwart particular data traces to become training data for machine learning. This way computational outputs manifest values that correspond to ethical virtues and that are socially accepted, appreciated, and sought-after like friendliness, literacy, truthfulness, positive emotionality, prosocial orientations, etc.

## 5.7 Systemic imperatives

A further apparent objection is to remark that beneficial machine learning applications, as supposed in this paper, stand in contradiction to systemic imperatives and goals of the economy. While making autonomous cars safer should result in having a competitive advantage, rendering recommendation systems on e-commerce platforms towards promoting sustainable, but more expensive products or on social media platforms towards less engaging content means that the platforms acquire fewer purchasers or users who can be influenced by online advertisements. However, the whole point of making something beneficial, in this case machine learning applications, is to overwrite systemic imperatives in case they have detrimental effects for particular individuals or society at large. Being successful according to the logic of a certain social system (Luhmann 1995) does not necessarily mean that this success is morally justified (Habermas 1987). This holds especially true with regard to the economy (Brand and Wissen 2017). On a related note, beneficial machine learning applications, which are traditionally used in areas like health or crisis response, satellite image interpretation, climate action, poverty reduction, wildlife preservation, and the like (Chui et al. 2018) do in many cases not follow systemic but moral imperatives. What is special in the case of the ideas presented here is that I do not propose to invent new machine learning application in hitherto undiscovered fields of society. Rather, I opt for reshaping applications that are already entrenched in areas that are structured by systemic imperatives and eventually aim at being profitable. In contexts that are purely dominated by monetary considerations, it is of course difficult to put the ideas into practice. Nevertheless, companies should at no point see the pursuit of profits as their only target. And as soon as they include social responsibility into their repertory of values, they can embrace the presented approach for beneficial machine learning.

# 6 Conclusion

In their classic book "Moral Machines", Wallach and Allen state: "The vision of learning systems developing naturally toward an ethical sensibility that values humans and human ethical concerns is an optimistic vision […]." (Wallach and Allen 2009, p. 110) This paper is a tangible proposal how this vision could be put into practice. It stresses the importance of "feeding" machine learning applications not with all relevant behavioral data that is available, but with a particular selection of it, namely with quality data. Following the typical big data approach and using all available data to train models can have detrimental effects. This can not only be shown by, for instance, pointing at various cases of algorithmic discrimination. It only recently got obvious when COVID-19 caused dramatic changes in online shopping and other digitally recorded behavior, so that its inclusion in training sets caused machine learning applications to malfunction, making manual interventions necessary (Heaven 2020). Thus, more rigorous mechanisms to filter training data sets have to be put in place, ensuring that, among others, only "good data" become training stimuli, meaning that digitally recorded behavior is classified and assessed along ethical criteria. Moral machine behavior is dependent on moral human behavior. Hence, both have to be linked.

Many machine learning applications acquire their "intelligence" by capturing aggregated human cognitive and behavioral abilities. Hitherto, those aggregations of recordings of human behavior are hardly presorted before becoming training stimuli for machine behavior. This paper is a plea to do so and thereby to achieve truly beneficial machine learning. Its arguments start at the assumption that a person's social background, dispositions, educational level, etc. shapes his or her way of using digital devices. In turn, those devices are able to track, infer, and measure a user's personal states or traits. Depending on these attributes, an ethical assessment of data traces the respective user produces, namely the data traces that provide the basis to the measurements itself as well as data traces that are situated in the same context as the measurement, can take place. Subsequently, this assessment enables to single out quality data that are representing behavior of individuals who are deemed to be especially competent, eligible, or morally versed for a particular task. This method for sampling out particular training stimuli stands in contrast to the *n = all* ideology. This serves to tailor training data in a way that machine behavior can correspond to values of the common good and become truly beneficial.

# 7 Acknowledgements

This research was supported by the Cluster of Excellence "Machine Learning – New Perspectives for Science" funded by the Deutsche Forschungsgemeinschaft (DFG, German Research Foundation) under Germany's Excellence Strategy – Reference Number EXC 2064/1 – Project ID 390727645. I would like to thank Sarah Fabi, Zeynep Akata, Ulrike von Luxburg, and Sebastian Bordt for very helpful comments on the manuscript.

# 8 Publication bibliography

Intelligence Technologies in the Near-Term. Available online at https://ainowinstitute.org/AI_Now_2016_Report.pdf, checked on 8/26/2020.

Čubranić-Dobrodolac, Marjana; Lipovac, Krsto; Čičević, Svetlana; Antić, Boris (2017): A Model for Traffic Accidents Prediction Based on Driver Personality Traits Assessment. In *PROMET* 29 (6), pp. 631–642.

Daly, Angela; Devitt, Kate S.; Mann, Monique (Eds.) (2019): Good Data. Amsterdam: Institute of Network Cultures.

Danaher, John (2018): Toward an Ethics of AI Assistants. An Initial Framework. In *Philos. Technol.* 31 (4), pp. 629–653.

Davidow, William; Malone, Michael S. (2020): Don't Regulate Artificial Intelligence: Starve It (Scientific American). Available online at https://blogs.scientificamerican.com/observations/dont-regulate-artificial-intelligence-starve-it/, checked on 5/8/2020.

Deleuze, Gilles (1992): Postscript on the Societies of Control. In *October* 59, pp. 3–7.

D'Mello, Sidney (2013): A selective meta-analysis on the relative incidence of discrete affective states during learning with technology. In *Journal of Educational Psychology* 105 (4), pp. 1082–1099.

D'Mello, Sidney; Kory, Jacqueline (2015): A Review and Meta-Analysis of Multimodal Affect Detection Systems. In *ACM Comput. Surv.* 47 (3), pp. 1–36.

Domingos, Pedro (2015): The Master Algorithm. How the Quest for the Ultimate Learning Machine Will Remake Our World. New York: Basic Books.

Dutta, Sanghamitra; Venkatesh, Praveen; Mardziel, Piotr; Datta, Anupam; Grover, Pulkit (2020): An Information-Theoretic Quantification of Discrimination with Exempt Features. AAAI Conference on Artificial Intelligence, pp. 1–28.

Dwork, Cynthia (2006): Differential Privacy. In David Hutchison, Takeo Kanade, Josef Kittler, Jon M. Kleinberg, Friedemann Mattern, John C. Mitchell et al. (Eds.): Automata, Languages and Programming. Berlin: Springer, pp. 1–12.

Dwork, Cynthia; Hardt, Moritz; Pitassi, Toniann; Reingold, Omer; Zemel, Richard (2011): Fairness Through Awareness. In *arXiv*, pp. 1–24.

Eady, T. A. (2019): Why Tesla's Fleet Miles Matter for Autonomous Driving (Medium). Available online at https://towardsdatascience.com/why-teslas-fleet-miles-matter-for-autonomous-driving-8e48503a462f, checked on 5/11/2020.

Engelmann, Severin; Chen, Mo; Fischer, Felix; Kao, Ching-yu; Grossklags, Jens (2019): Clear Sanctions, Vague Rewards: How China's Social Credit System Currently Defines "Good" and "Bad" Behavior. In *Proceedings of the Conference on Fairness, Accountability, and Transparency - FAT* '19*, pp. 69–78.

Erikson, Erik H. (1980): Identity and the life cycle. New York: W.W. Norton.
23